\documentclass[aps,reprint,superscriptaddress,showpacs,nofootinbib]{revtex4-2}

\usepackage{hyperref}
\usepackage{amsmath}
\usepackage{amssymb}
\usepackage{graphicx}
\usepackage{color}
\usepackage{subfig}
\usepackage{bm}
\usepackage{bigints}
\usepackage[usenames,dvipsnames,svgnames,table]{xcolor}
\usepackage{pifont}
\usepackage{aas_macros}
\usepackage{physics}
\usepackage{ulem}

\usepackage[labelfont=bf]{caption}

\hypersetup{
    colorlinks=true,
    linkcolor=Blue,
    filecolor=Blue,
    urlcolor=MidnightBlue,
    citecolor=Blue,
   pdftitle={Cooling age}
}

\begin{document}

\title{
Einstein crystals in Snyder and Snyder-de Sitter noncommutative backgrounds}

\author{Anna Pacho\l}
\email[E-mail: ]{anna.pachol@usn.no}
\affiliation{ Department of Microsystems, University of South-Eastern Norway, Campus Vestfold, Norway}
\affiliation{Theoretical Sciences Visiting Program, Okinawa Institute of Science and Technology Graduate University, Onna, 904-0495, Japan}

\author{Aneta Wojnar}
\thanks{Corresponding author}
\email[E-mail: ]{aneta.wojnar@uwr.edu.pl}
\affiliation{Institute of Theoretical Physics, University of Wroc\l aw, pl. Maxa Borna 9, 50-206 Wroc\l aw, Poland}

\begin{abstract}
We investigate the behaviour of Einstein crystals in noncommutative backgrounds described by the Snyder and Snyder–de Sitter models. Possible novel effects, which may arise in realistic systems such as diamond crystals, are analysed within a thermodynamical framework. We show that noncommutativity influences the key thermodynamic quantities, including internal energy and specific heat. These corrections can be directly related to modifications of the underlying uncertainty relations, of the generalized uncertainty principle (GUP) and generalized extended uncertainty principle (GEUP) types.
\end{abstract}

\maketitle

\section{introduction}
The conventional quantum gravity research focused for a long time on the search of quantum gravitational effects at the Planck-scale (very high energies $10^{19}$ GeV or correspondingly very small length scales $10^{-35}$ m). Consequently, direct experimental access to such effects has been considered unattainable. Recently, however, substantial progress in materials science and quantum engineering allowed for reaching unprecedented scales in controlling the size and mass of quantum systems in the laboratory settings \cite{pedalino2026probing}.
These developments open a new way to explore low-energy regimes, in which gravity or possible extensions of quantum theory can be investigated in complement to accelerator-based, interferometric, and cosmological approaches.
For example, there are various proposals where massive quantum systems are considered as the interface for quantum and gravitational interactions, see e.g. the recent review \cite{Bose:2024nhv}.

These advances suggest that the intersection of quantum material science and gravitational physics may provide a realistic pathway toward experimentally probing the quantum nature of gravity.

In the search for quantum and gravitational interactions at the low energy levels one can consider extensions or modifications of quantum theory. For example modified (non-canonical) commutation relations of quantum mechanical phase space may lead to the appearance of new nonzero uncertainty in position measurements in addition to position and momenta uncertainty relations due to the quantization of position observables (noncommutativity). Such generalised Heisenberg relations and uncertainty principle (GUP) impact the Schr\"odinger equation and the energy states get additional terms \cite{Chang:2001kn}.

For example, there have been developments in the direct table-top experiments, considering the low-energy mechanical oscillators, designed to probe the effects arising from the corrections introduced in the Heisenberg relations and GUP. In \cite{bawaj2015probing}, the free evolution of micro-and nano-oscillators has been analysed. Moreover, by exploring crystal materials and phonon mechanical systems one can try to detect frequency perturbations in mechanical resonators (by using quantum resonances and phonon sensitivity to gravitational fields), see e.g 
\cite{campbell2023improved}.
Crystals and other crystalline materials offer an interesting venue for probing quantum aspects of gravity or possible extensions of quantum theory.  
Their well characterized mechanical, optical, and elastic properties allow for such investigations  with physical predictions which may be possible to be tested in the future.

In this work we shall focus on the modifications of quantum mechanical phase space, motivated by the mathematical framework of the noncommutative geometry (NCG) and quantum space(-time)s. By choosing a specific model of quantum (NC) space, one naturally obtains the deformation (or modification) in the canonical quantum mechanical phase space. This in turn leads to the modifications in the uncertainty principles, for example the Snyder model leading to GUP, while Snyder-de Sitter (SdS) to generalized extended uncertainty principle (GEUP), see e.g. \cite{Pachol:2024hiz}. 

In the first part of this paper (Sec. 2 and 3), we will rely on the Snyder model (the first proposition of noncommutative Lorentz invariant space-time) \cite{snyder1947quantized}.  We consider it in the non-relativistic and 1 dimensional version with the commutation relations between the position $\hat{x}$ and
momentum $\hat{p}$ operators described by
\begin{equation}\label{xp}
\left[ \hat{x},\hat{p}\right] =i\hbar \left( 1+\theta \hat{p}^{2}\right)
\end{equation}
where 
\begin{equation}
    \label{delta}
    \theta=\zeta(3\xi-\frac{1}{2})
\end{equation}
includes the noncommutativity parameter $\zeta$ 
of dimension [$\frac{L^{2}}{\hbar ^{2}}]$ that sets the scale of the modification. In the quantum gravity context $L$ - length is usually associated with the Planck length $L_{P}$ but it could be considered as length of any order just below currently measurable size  $\sim 10^{-19}$ m and interpreted as an extension of QM. The parameter $\xi$ encodes possible various realizations of the Snyder model (e.g. $\xi=0;\, 1/2; \, 1/6$)  \cite{Battisti:2010sr,Pachol:2023tqa}.
\eqref{xp} 
provides the first order of modifications (in noncommutativity parameter $\zeta$) of the Heisenberg canonical commutation relations. The undeformed limit (the usual quantum theory) is recovered for $\zeta\to 0$, i.e. $\theta\to 0$.
The relation \eqref{xp} includes
the most general realization of the Snyder model \cite{Battisti:2010sr}, where for different values of
parameter $\xi $ we can recover various forms of these commutation
relations commonly used in the literature, for example:\\
\textbullet $\qquad\xi=\frac{1}{2}$ (i.e. $\theta =\zeta$) reduces
to the original Snyder realization \\
\textbullet $\qquad\xi=0$ (i.e. $\theta =-\frac{1}{2}\zeta$) is related to the "Maggiore realization" (in a compact form expressed as a square-root).\\
\textbullet $\qquad\xi=\frac{1}{6}$ (i.e. $\theta = 0$) Weyl realization which introduces no modification to \eqref{xp}. \\
It is worth to point out that similar relation as in \eqref{xp} is often considered in purely phenomenological models, such as Generalized Uncertainty Principle (GUP) see e.g. \cite{Bosso:2023aht} for a recent overview of various approaches. In this context, the first type of realization ($\xi=1/2$), leads to the standard quadratic GUP (QGUP) relations, while the second one ($\xi=0$) is related to the so-called higher order type GUP, see e.g.  \cite{Maggiore:1993zu,Segreto:2022clx}. 

One can also consider the case of modified quantum phase space \eqref{xp} with the negative coupling parameter $\zeta$, then we call such model anti-Snyder, see e.g. \cite{Mignemi:2011gr,Mignemi:2011wh} where  the non-relativistic Snyder and anti-Snyder models were considered.
In this case 
the anti-Snyder model, would include $$
\left[ \hat{x},\hat{p}\right] =i\hbar \left( 1-\lambda(3\xi-\frac{1}{2})\hat{p}^{2}\right).$$ For simplicity of the notation we shall use $\zeta$ and \eqref{xp} and admit both positive and negative values of $\zeta$, i.e. taking $\zeta=-\lambda$ to obtain the anti-Snyder model.

In the second part of this work (Sec. 4), we will consider the generalizations of Snyder model to the curved spaces resulting in Snyder-de Sitter (SdS) NC background with relations reducing in 1 dimension to
$$    \lbrack \hat{x},\hat{p}]=i\hbar \left( 1+\alpha\hat{x}^2+\zeta \hat{p}^2+\sqrt{\alpha \zeta} (\hat{x}\hat{p}+\hat{p}\hat{x})\right). 
$$
These relations lead to a different type of the modification of the uncertainty principle, often referred to as GEUP/EGUP, which admit both fundamental parameters $\zeta$ and $\alpha$ which may be related to the Planck length and cosmological constant, respectively or may be related to the possible NC corrections appearing in quantum theory.
In its 3 dimensional non-relativistic version, when both of the
coupling constants $\zeta,\alpha$ are positive, SdS model corresponds to the Snyder model
restricted to a three-dimensional sphere. For the negative coupling constants $\zeta,\alpha$ the non-relativistic model is obtained by considering the anti-Snyder model on a pseudosphere, this case was called anti-Snyder-de Sitter (aSdS) model in \cite{Mignemi:2011gr,Mignemi:2011wh}
and we use this name here.

In this paper, we investigate the effects of the GUP and GEUP on crystal systems within a thermodynamical framework. 
We consider the Einstein's model of the thermal properties of crystals at low temperatures.
In this approach, atoms in the crystal are modeled as independent, thermally excited harmonic oscillators, with interactions between atomic vibrations neglected.
This simplified model is commonly used in solid-state physics allowing to understand lattice vibrations, thermal properties such as heat capacity, and the quantum nature of these vibrations, which are quantized as phonons. 

Indications that NC geometry may influence crystal properties have already appeared in our previous work \cite{Pachol:2023tqa} and in \cite{riasat2023effect}.
Here we focus on the GUP (and later GEUP)-induced modifications to the energy spectrum of a one-dimensional harmonic oscillator, following \cite{chang2002effect} (and \cite{Mignemi:2011wh} respectively), and derive the corresponding partition function, including NC corrections up to the first order in the deformation parameter. This allows us, in particular, to obtain indicative constraints on the NC parameter from the requirement of positivity of the partition function.
We then compute the internal energy and specific heat, and discuss the physical implications of the resulting NC corrections.
\section{Partition Function of a 1D oscillator in Snyder (GUP) background}
Starting with the Hamiltonian of a 1D harmonic oscillator $$\hat{H}=\frac{1}{2}\mu \omega ^{2}\hat{x}^{2}+\frac{1}{2\mu }\hat{p}^{2}$$ we want to incorporate the modifications arising from noncommutativity \eqref{xp} (or a generalized uncertainty principle (GUP) corresponding to \eqref{xp}). Our derivation bases on the solution to the Schr\"odinger equation
 obtained in the presence of the
modified quantum mechanical phase space relations found in \cite{Chang:2001kn}. Therein, the exact 1D harmonic oscillator's energy, in the presence of quantum phase space relations as in \eqref{xp}, was obtained as
\begin{eqnarray}
E_{n} &=&\hbar\omega \left[ \left(n + \frac{1}{2}\right)\sqrt{1 + A^2 } + \left(n^2+n+\frac{1}{2}\right)A\right]\nonumber\\
n&=&0,1,2\ldots,
\label{En1D_27short}
\end{eqnarray} 
where $A=\frac{1}{2}\mu\hbar\omega\theta$ is linear in $\zeta$ - noncommutativity parameter ($\theta$ includes $\zeta$, see \eqref{delta}) and $\mu$ is the mass of the oscillator and $\omega$ its frequency.
Since the term $A^2$ is considered to be small
we use the approximation $\sqrt{1+A^2}\approx 1+A^2/2$ and neglect the $A^2$ order terms\footnote{Note that this is consistent with the relations \eqref{xp} as already in  \eqref{xp} we only consider the terms up to and including the first order in the noncommutatitvity parameter $\zeta$.}. Then the energy becomes
\begin{eqnarray}\label{En_approx}
E_{n} &=&
\frac{\hbar\omega}{2}(1+A)  +\hbar\omega\left[n(1+A) + n^2A\right].
\end{eqnarray} 
For the undeformed limit ($\zeta\to 0$) $A\to 0$ the energy eigenvalues of a simple harmonic oscillator 
take the usual quantized, but undeformed, form
\begin{eqnarray}
(E_{n})_{undef} &=&\hbar\omega \left(n + \frac{1}{2}\right).
\label{En_undef}
\end{eqnarray} 
However, in the presence of the NC corrections (which could be interpreted as new physical effects arising from NC extension of quantum theory or as modifications arising from quantum gravity), we can see that the energy level with $n=0$, which is the ground state, has the modified (deformed) energy
\begin{equation}\label{E_0zdef}
    E_{0} = 
\frac{\hbar\omega}{2} (1+A)= \frac{\hbar\omega}{2}
(1+\frac{\zeta}{2}\mu\hbar\omega(3\xi-\frac{1}{2})).
\end{equation}

Let us now consider a partition function of a 1D harmonic oscillator in the noncommutative case with the energy eigenvalues as in \eqref{En_approx}. 
Working with the standard notation, that is, $\beta=\frac{1}{k_BT}$, the partition function of the considered case becomes
\begin{align}
    Z&=\sum_{n=0}^\infty e^{-\frac{E_n}{k_BT}}
    =e^{-\frac{1}{2}\beta\hbar\omega (1+A)}\sum_{n=0}^\infty e^{-\beta\hbar\omega \left[n(1+A) + n^2A\right]}\nonumber\\
&
\label{sum}
    =e^{-\beta E_0}\sum_{n=0}^\infty e^{-\beta\hbar\omega \left[n(1+A) + n^2A\right]}.
\end{align}
One can show that the above series is convergent for $A >-\frac{1}{2}$. In the undeformed limit $\zeta\to 0$ (i.e. $A\to 0$, or when the Weyl realization is chosen for the Snyder model, $\xi=\frac{1}{6}$ giving $A=0$), we recover the well-known canonical case $Z_{undef}$ consisting of the geometric series
\begin{equation}
    Z_{undef}= e^{-\beta (E_0)_{undef}}
\sum_{n=0}^\infty \Big( e^{-\hbar\omega\beta }\Big)^n= \frac{e^{-\frac{\beta \hbar\omega}{2}}}{1-e^{-\beta\hbar\omega}}.\label{cl_HO}
\end{equation}
However, in the noncommutative case such summation is more complicated. Considering the small NC parameter $\zeta$ (therefore, small $A$),
the sum \eqref{sum} can be rewritten as{\footnotesize
\begin{align}
 &Z=e^{-\beta E_0}\sum_{n=0}^\infty e^{-\beta\hbar\omega n} e^{-\beta\hbar\omega (n+n^2)A}\\
&
=e^{-\beta E_0}\Big(\sum_{n=0}^\infty e^{-\beta\hbar\omega n}- \beta\hbar\omega A \sum_{n=0}^\infty ne^{-\beta\hbar\omega n}- \beta\hbar\omega A \sum_{n=0}^\infty n^2 e^{-\beta\hbar\omega n}\Big) \nonumber
\end{align}}
where we expanded the last exponent in $A$. Thanks to this, each of these terms can now be summed over, providing
\footnotesize{
\begin{equation}
     Z=e^{-\beta E_0}
     \Big(\frac{e^{\beta\hbar\omega}}{e^{\beta\hbar\omega}-1}-\beta\hbar\omega\frac{Ae^{\beta\hbar\omega}}{(e^{\beta\hbar\omega}-1)^2}-\beta\hbar\omega\frac{Ae^{\beta\hbar\omega}(e^{\beta\hbar\omega}+1) }{(e^{\beta\hbar\omega}-1)^3}
     \Big).
\end{equation}}\normalsize
 Now, let us use the expression \eqref{E_0zdef} for $E_0$  (we emphasise that $E_0$ is modified and contains the correction in $A$) in the above formula and do a similar expansion in $A$ as before (i.e. up to the linear order in $\zeta$). Thanks to this we get
\begin{equation}
     Z=e^{-\frac{\beta\hbar\omega}{2}}\Big(\frac{e^{\beta\hbar\omega}}{e^{\beta\hbar\omega}-1}-A\beta\hbar\omega \frac{e^{\beta\hbar\omega}(e^{\beta\hbar\omega}+1)^2}{2(e^{\beta\hbar\omega}-1)^3}
     \Big)
\end{equation}
where the correction to the undeformed partition function $Z_{undef}$ is clear, cf. \eqref{cl_HO}.

Now, since we need to take the logarithm of this expression, $Z$ must always be positive, which leads to the condition that
\begin{equation}\label{signA}
   \Big(1-A\beta\hbar\omega \frac{(e^{\beta\hbar\omega}+1)^2}{2(e^{\beta\hbar\omega}-1)^2}
     \Big)>0 .
\end{equation}
This condition can be used to obtain the constraints on the NC parameter $\zeta$, full analysis can be found in Appendix \ref{ln_ineq}. To obtain the values for the constraints, we consider a diamond crystal\footnote{A diamond crystal is well described by the Einstein model in low temperatures. We have taken the energy of a {quantum} as $\hbar\omega=0.19 \text{eV}\ \sim 0.30441356046\times 10^{-19}$J while the effective mass of the oscillator is $9.96\times10^{-27}$kg.}. 

Considering the Maggiore realization of the Snyder model, that is, $\xi=0$, the analysis of \eqref{signA} provides the following constraints (in [J$\cdot$ kg]):
\begin{equation}
    \begin{cases}
	\zeta>-1.2\times10^{43}T, &  k_B T<<\hbar\omega\\
    \zeta>-1.46\times10^{49}\frac{1}{T}, &   k_B T>>\hbar\omega.
		 \end{cases}
\end{equation}
Taking into account the convergence condition discussed before, we have also the upper bound $\zeta<6.6\times 10^{45}$ in this case.

On the other hand, for the case $\xi=\frac{1}{2}$ (the original Snyder realization), one obtains
\begin{equation}\label{part_val12}
    \begin{cases}
	\zeta<5.9\times 10^{42}T, &   k_B T<<\hbar\omega\\
   \zeta<7.27\times10^{48}\frac{1}{T}, &   k_B T>>\hbar\omega.
		 \end{cases}
\end{equation}
Considering the convergence condition, one deals with the lower bound $\zeta>-3.3\times 10^{45}$.

As expected, stronger constraints are obtained in the very low-temperature regime, where the Einstein model more accurately describes crystalline solids.

\section{Internal energy and Heat Capacity of 3-dim solids}
Since we want to consider a 3 dimensional (3D) solid at low temperatures modeled by the Einstein crystal,  let us assume that each oscillator has the same frequency $\omega$, so the total energy of the 3D crystal with $N$ oscillators can be expressed as\footnote{One can consider the general $D$ dimensional case, basing on the results for the energy eigenvalues obtained in \cite{chang2002effect,Chang:2001kn} and we postpone this case for the future.}
 \begin{equation}
    E_{tot} = \sum^{3N}_{i=0} (E_n)_i=3N  E_n,
\end{equation}
where
$(E_n)_i=\hbar\omega \left[ \left(n + \frac{1}{2}\right)\sqrt{1 + A^2 } + \left(n^2+n+\frac{1}{2}\right)A\right]$ is identical for each oscillator, cf. \eqref{En1D_27short}.
The ground state energy of the solid is given by
{\small \begin{equation}
    E_{tot} = \sum^{3N}_{i=0} E_0 = 3NE_0.
\end{equation}}
The internal (average) energy of the ensemble of oscillators based on the partition function calculated above can be written as
{\small
\begin{equation}\label{Uf}
    U= - 3N \frac{\partial\, \ln Z}{\partial\beta}
    = \frac{3N\hbar\omega}{2}\coth
   \left(\frac{\hbar\omega}{2 k_B T}\right)-A\frac{3N\hbar\omega}{2}f(\omega,T),
\end{equation}
}
where, for convenience, we have introduced
\begin{equation}
    f(\omega,T)=\frac{ \left(e^{\hbar\beta \omega }+1\right) }{\left(e^{\hbar \beta \omega }-1\right) ^{3}}\left(1-e^{2(\hbar\beta \omega
)}+4\hbar\beta \omega e^{\hbar\beta \omega }\right)
\end{equation}
and we have already performed an expansion in $A$. As expected,  \eqref{Uf} reduces to the known (undeformed) case for $A\rightarrow0$.
It can be easily shown that  \eqref{Uf} can be also rewritten in terms of trigonometric expressions as
{\footnotesize 
\begin{align}\label{u_total}
    U    =& 3N\frac{{\hbar\omega}}{2}\coth
   \left(\frac{\hbar\omega}{2 k_B T}\right)+\\ 
   +&3AN\frac{{\hbar\omega}}{8}  
   \sinh\left(\frac{\hbar\omega}{k_B T}\right)\csch^4\left(\frac{\hbar\omega}{2 k_BT}\right)
   \left(\sinh \left(\frac{\hbar\omega}{k_B T}\right)-\frac{2\hbar\omega}{k_BT}\right).
   \nonumber
\end{align}}
To consider the cases of interest (low- and high-temperatures), it will be easier to rewrite the function $f(\omega,T)$ in terms of a dimensionless variable $x=\frac{\hbar\omega}{k_B T}$
\begin{equation}
   f(x)=\frac{ \left(e^{x }+1\right) }{\left(e^{x }-1\right) ^{3}}\left(1-e^{2x}+4x e^{x }\right).
\end{equation}
Considering the asymptotical behavior for both $x\gg 1$ (i.e. low temperatures $\hbar\omega\gg k_B T$) as well as $x\ll 1$ (i.e. high temperatures $\hbar\omega\ll k_B T$) one obtains
\begin{equation}
   f(x)\approx
    \begin{cases}
	  -1+4xe^{-x}, & \text{if $x\gg 1$ }\\
     \frac{4}{x^2}      , & \text{if $x\ll 1$ },
		 \end{cases}
\end{equation}
where we have used $\coth(x)\approx 1/x$ for $x\ll 1$ and $\coth(x)\approx 1$ for $x\gg1$.
Therefore, the internal energy is given by
\begin{equation}\label{Uf2}
    U= 
     \begin{cases}
	  \frac{3N}{2} x k_B T(1+A(1-4xe^{-x}), & \text{if $x\gg 1$ }\\
     3N k_B T(1-\frac{2A}{x})    , & \text{if $x\ll 1$ } 
		 \end{cases}
\end{equation}
which, after returning to the old notation with $x=\frac{\hbar\omega}{k_B T}$, turns into
{\small 
\begin{equation}\label{uf}
    U=
    \begin{cases}
	\frac{3}{2}N \hbar\omega  \left(
    1+A\left( 1-4\frac{\hbar \omega}{k_B T} e^{-\frac{\hbar \omega
   }{k_B T}}\right)\right), & \text{if $k_B T \ll \hbar \omega$ }\\
          3Nk_BT \left(1-2A\frac{ k_B T}{\hbar\omega}\right) , & \text{if $k_B T \gg \hbar \omega$ }.
		 \end{cases}
\end{equation}
}
It is now straightforward to obtain the heat capacity of the considered system. In general case from \eqref{u_total} one obtains
{\small 
\begin{align}\label{specific0}
    C_V=&\frac{dU}{dT}=\frac{3 N (\hbar\omega)^2 }{4 k_B
   T^2\text{sinh}^2\left(\frac{\hbar\omega}{2 k_B T}\right)}\\
   -&A\frac{3N \left(\frac{\hbar\omega}{k_BT}\right)^2  \left[\hbar \omega  (\cosh
   \left(\frac{\hbar\omega}{k_BT}\right)+2)-2 k_B T \sinh \left(\frac{\hbar\omega}{k_BT}\right)\right]}{4 T\text{sinh}^4\left(\frac{\hbar\omega}{2k_BT}\right) }.\nonumber
\end{align}
}
Considering the asymptotical behavior for both  low and  high temperatures 
one obtains
{\small 
\begin{equation}\label{limiting}
    C_V=
    \begin{cases}
	3 N  k_B e^{-\frac{\hbar\omega}{k_B T}}
\Big(\frac{\hbar\omega}{k_BT}\Big)^2 (1-2\Big(\frac{\hbar\omega}{k_BT}\Big)A), & k_B T \ll \hbar \omega,\\
    3Nk_B(1-4A \Big(\frac{k_B T}{\hbar \omega} \Big)), & k_B T \gg \hbar \omega .
		 \end{cases}
\end{equation}}
\normalsize
To find the permitted values of the parameter $\zeta$ in different temperature/energy regimes, we first look at the range for which the rescaled specific heat (i.e. $\tilde C_V = C_V /(3N k_B )$) is always positive but smaller than 1. For this purpose we analyse the behavior of rescaled heat capacity at different temperature ranges, in terms of a new dimensionless variable $y=\frac{1}{x}=\frac{k_B T}{\hbar\omega}$ (as the reciprocal of $x$ - gives the standard relation with the temperature increasing toward the right). From \eqref{specific0} we obtain
\begin{align}\label{CVplot1}
    \frac{C_V}{(3Nk_B)}=&\frac{1}{4y^2\text{sinh}^2\left(\frac{1}{2y}\right)}\\
   -&\frac{A  \left[\frac{1}{y}  (\cosh
   \left(\frac{1}{y}\right)+2)-2  \sinh \left(\frac{1}{y}\right)\right]}{4y^2\text{sinh}^4\left(\frac{1}{2y}\right) }.\nonumber
\end{align}
We note that there exists a value of $A_{crit}$ (hence the value of NC parameter $\zeta_{crit}$) for which the specific heat $C_V$ can vanish (and, of course, $C_V$ must never be negative for solids). This gives:
\begin{equation}
    A_{crit}=\frac{\text{sinh}^2\left(\frac{1}{2y}\right) }{\left[\frac{1}{y}  (\cosh
   \left(\frac{1}{y}\right)+2)-2\sinh \left(\frac{1}{y}\right)\right]}.\nonumber
\end{equation}
With $A=\frac{1}{2}\mu\hbar\omega\zeta(3\xi-\frac{1}{2})$ and the values for the diamond crystal as before (see footnote 2) this gives:
\begin{equation}\label{zeta_crit}
    \zeta_{crit}=\frac{6.59\times 10^{45}\text{sinh}^2\left(\frac{1}{2y}\right) }{(3\xi-\frac{1}{2})\left[\frac{1}{y}  (\cosh
   \left(\frac{1}{y}\right)+2)-2\sinh \left(\frac{1}{y}\right)\right]}.
\end{equation}
We perform the analysis of the rescaled heat capacity at different temperatures for both realizations, $\xi=0$ and $\xi=\frac{1}{2}$, of the Snyder model in the next section and the results are presented in Fig. \ref{zeta1} and \ref{zeta2}, respectively. 

\subsubsection*{Realization $\xi=0$}
Taking $\xi = 0$ in \eqref{zeta_crit} the following critical value of $\zeta_{\text{crit}}$, for which $C_V = 0$, is obtained
\begin{align}
\zeta_\text{crit}&\approx -\frac{1.319\times 10^{46}\text{sinh}^2\left(\frac{1}{2y}\right) }{\left[\frac{1}{y} (\cosh
\left(\frac{1}{y}\right)+2)-2\sinh \left(\frac{1}{y}\right)\right]}.
\label{crit0}
\end{align}
and it is valid for all temperature ranges\footnote{For example, considering diamond crystal and $T~\sim 200$ K we obtain $y_{200K}= 9.0708\times10^{-2}$ leading to (for $\xi=0$ realization) $\zeta_{crit-200K}=-7.3\times10^{44}$.}.

\begin{figure}[h!]
     \includegraphics[scale=0.43]{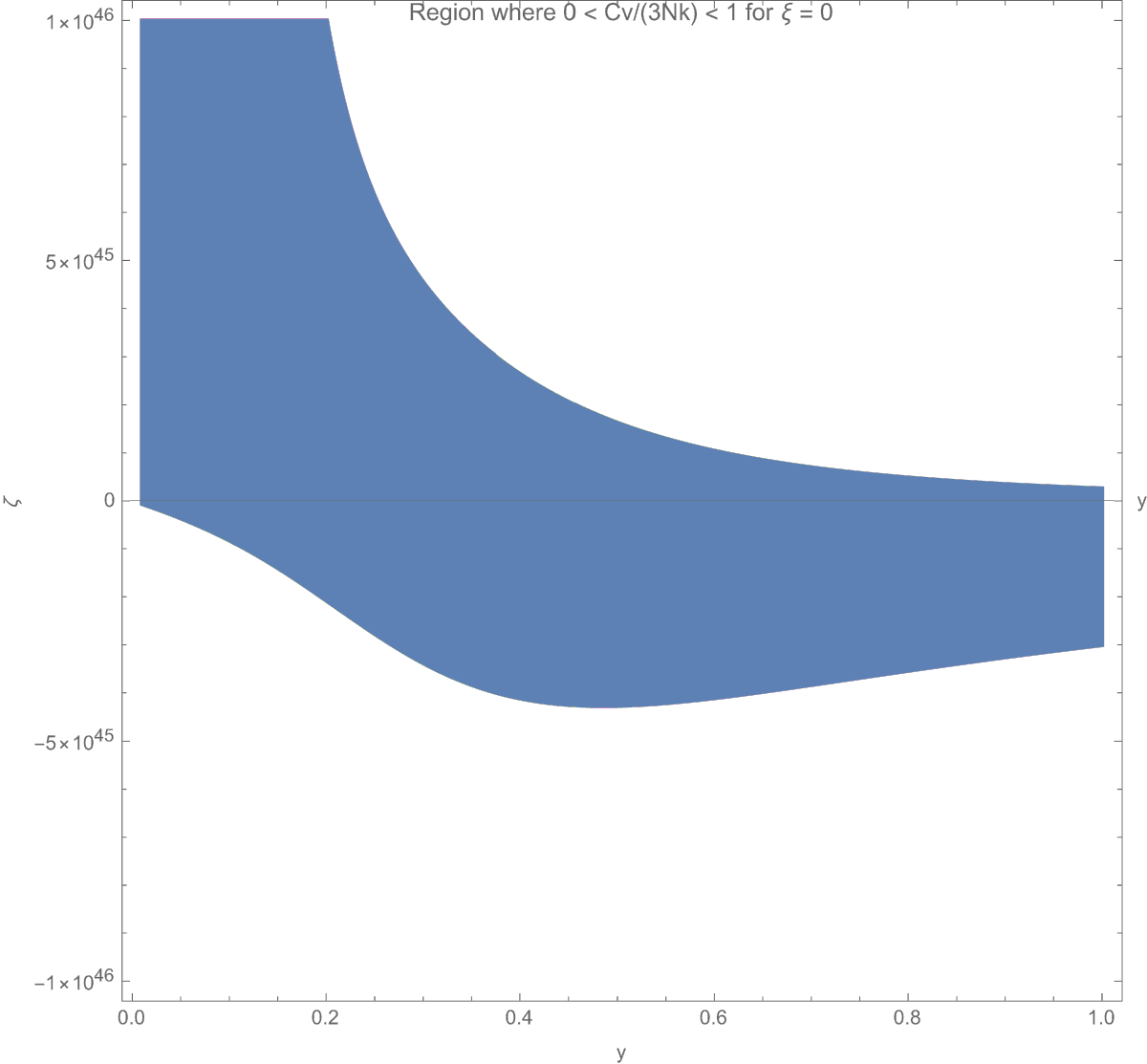}
     \caption{Allowed values of the parameter $\zeta$ in different temperature/energy range for which the values of the rescaled specific heat 
     lie in the range $0<C_V/(3Nk_B)<1$, that is, $\tilde C_V$ is always positive {but smaller than  1} (shaded regions) for the case $\xi=0$. }
     \label{zeta1}
\end{figure}

Note that at low temperatures the specific heat is a rapidly decreasing function, see \eqref{limiting}. The NC term accelerates this decrease and can even drive $C_V$ to negative values. Therefore, the critical value $\zeta_{\text{crit}}$ \eqref{crit0} prevents this behavior, especially at low temperatures (small values of $y$). Although the Einstein model is not suitable for the high-temperature regime, it is still necessary to recover the proper behavior of the $C_V$ curve (see the more detailed discussion below). Therefore, imposing $\tilde C_V= C_V/(3Nk_B) = 1$ provides further constraints on the allowed values of the NC parameter $\zeta$. Taking this into account, the region of feasible values of the parameter $\zeta$ for different temperature ranges in the realization $\xi = 0$ is shown in Fig.~\ref{zeta1}. 

\begin{figure}[h]
     \includegraphics[scale=0.55]{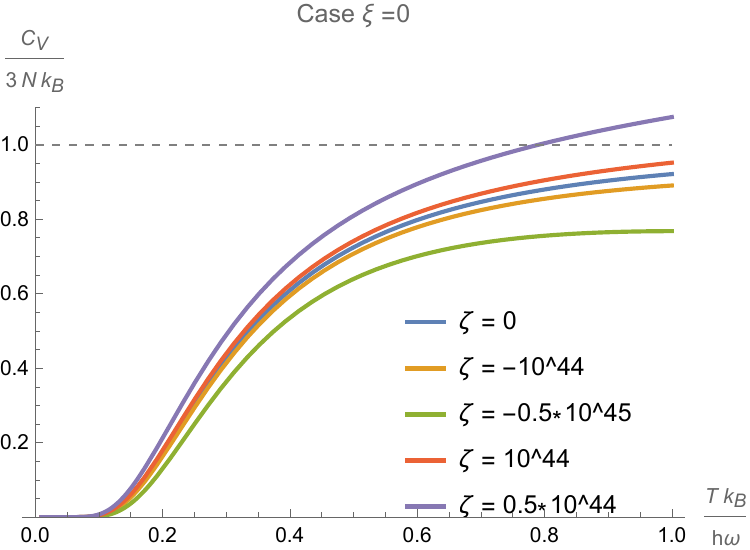}
     \caption{Specific heat \eqref{specific0} for different allowed values of the parameter $\zeta$ for $\xi=0$. It agrees with the convergence condition $\zeta<6.6\times 10^{45}$.}
     \label{cv1}
\end{figure}

At low temperatures the positive values of $\zeta$ are unbounded, since the correction term is a rapidly decreasing function (see also the low-temperature case \eqref{limiting}). However, increasing the parameter eventually drives the system into an {unphysical regime, as can be seen in Fig.~\ref{cv1}}, which shows the specific heat \eqref{CVplot1} for different values of $\zeta$.
This is because the Einstein model does not accurately describe the behavior of, for example, a diamond crystal at high temperatures. As can be seen from \eqref{limiting}, the correction term cannot drive the specific heat far from unity. Therefore, taking into account both low- and high-temperature limits, the analysis of Fig.~\ref{cv1} suggests that the parameter $\zeta$ must lie within the range $\zeta \approx \pm 10^{44}$.\\

Let us note that the "lower {boundary" curve in Fig.~\ref{zeta1} corresponds to the case for which $\tilde C_V = 0$ over the entire temperature range. This implies that, in principle, the specific heat could vanish even at high temperatures. Furthermore, there exists a minimum of \eqref{zeta_crit} around $y \approx 0.48$ ({which, however corresponds to $T\sim 1000$ K for the diamond, the temperature too high for crystals to be described by the Einstein model), for which $\zeta \approx -4.3 \times 10^{45}$. Although the Einstein model is not suitable for describing crystals at high temperatures, the emergence of different bounds in different temperature regimes appears to be a generic feature of both classical and quantum extensions of GR, as indicated by our previous studies \cite{Pachol:2023tqa,Kozak:2021vbm,Wojnar:2024xdy}.

On the other hand, the condition $\tilde C_V = 1$ for the allowed values of the parameter $\zeta$ is represented by the "upper {boundary}" curve in Fig.~\ref{zeta1}. One encounters an analogous situation to that of the vanishing specific heat: while the normalized specific heat approaches $\tilde C_V \rightarrow 1$ at high temperatures, this limit may also be reached already at values as low as $y \approx 0.2$ ({corresponding to temperatures $T\sim 440K$, for which the Einstein model is a good approximation for diamond}), giving rise to a nonphysical effect.

\subsubsection*{Realization $\xi=1/2$}
The realization $\xi=\frac{1}{2}$ in \eqref{zeta_crit} gives the following critical values of $\zeta_{crit}$'s for which $\tilde C_V=0$:
\begin{align}
   \zeta_\text{crit}&\approx 
     \frac{6.596\times 10^{45}\text{sinh}^2\left(\frac{1}{2y}\right) }{\left[\frac{1}{y}  (\cosh
   \left(\frac{1}{y}\right)+2)-2\sinh \left(\frac{1}{y}\right)\right]}.
\end{align}
The region of feasible values of the parameter $\zeta$ for different temperature ranges for this realization is plotted in Fig. \ref{zeta2}.
\begin{figure}[h!]
     \includegraphics[scale=0.43]{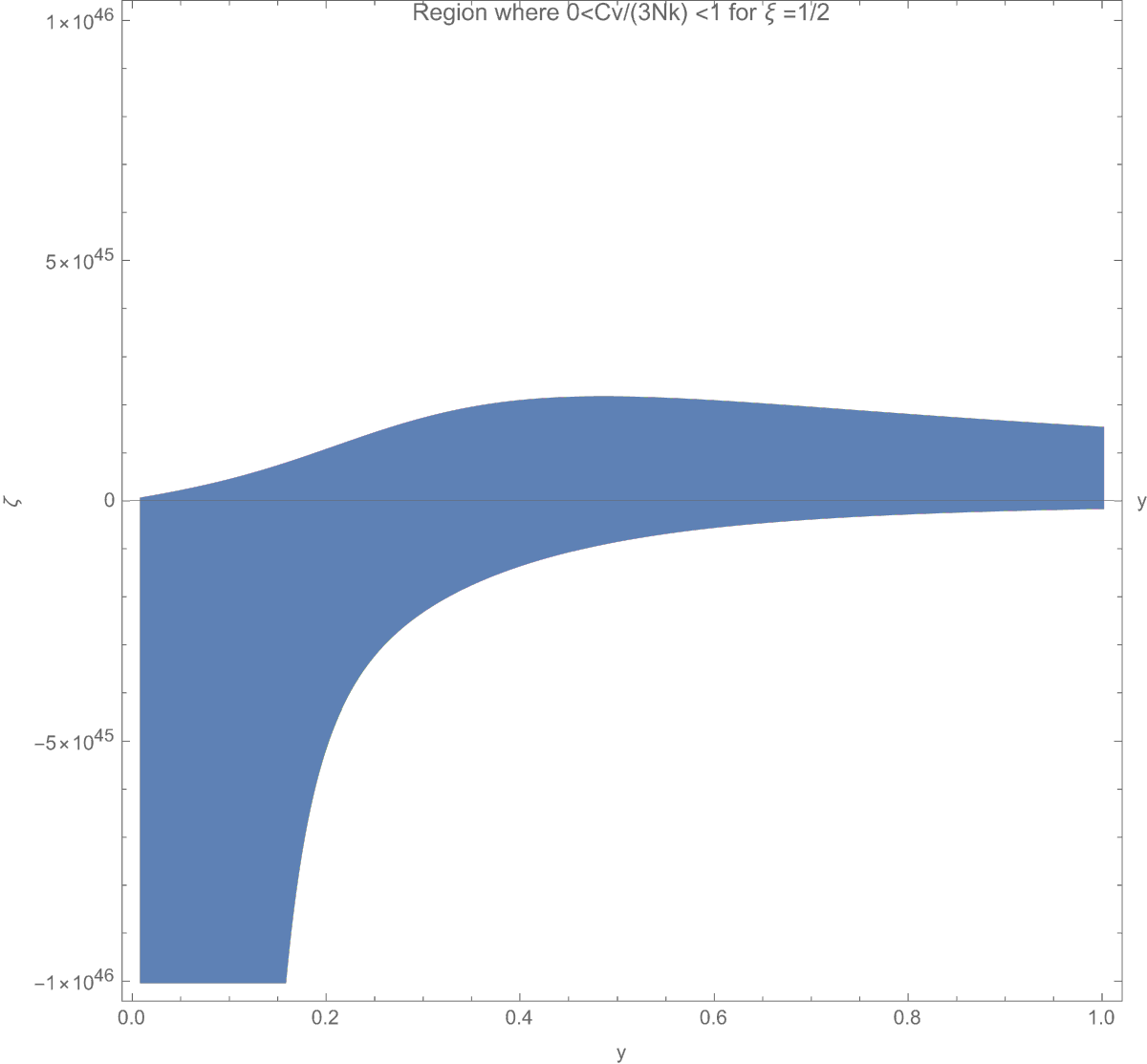}
     \caption{Allowed values of the parameter $\zeta$ in different temperature/energy range for which the specific heat \eqref{specific0}  lie in the range $0<C_V/(3Nk_B)<1$, that is, $\tilde C_V$ is always positive but smaller than 1 (shaded regions) for the case $\xi=\frac{1}{2}$.}
     \label{zeta2}
\end{figure}

Note that for low temperatures, the negative values of $\zeta$ are unbounded because the correction term is a rapidly decreasing function (see also \eqref{limiting}). However, increasing the parameter takes us to the unphysical region of the values of the specific heat as can be seen on the Fig. \ref{cv2}, representing the specific heat \eqref{specific0} for different values of $\zeta$. 

\begin{figure}[h]
     \includegraphics[scale=0.55]{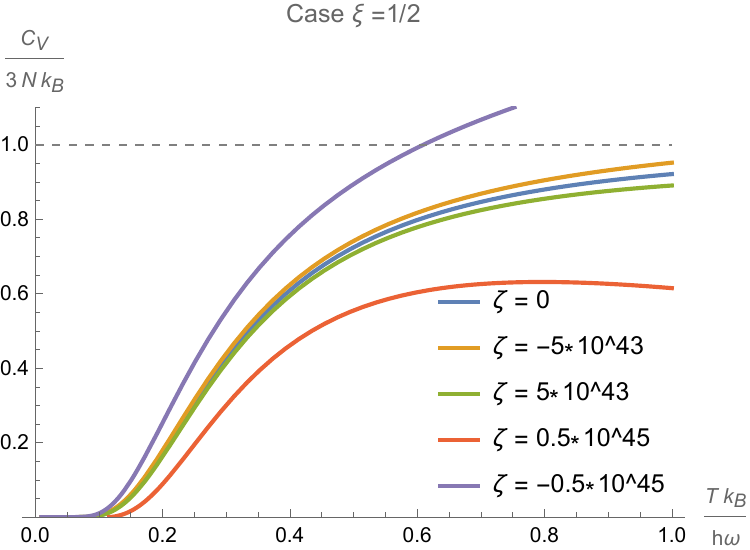}
     \caption{Specific heat \eqref{specific0} for different allowed values of the parameter $\zeta$ for $\xi=\frac{1}{2}$. It agrees with the convergence condition $\zeta>-3.3\times 10^{45}$.}
     \label{cv2}
\end{figure}
On the other hand, although the Einstein model does not describe well the behaviour of (for example) diamond crystal in high temperatures, we see from \eqref{limiting} that the correcting term cannot take us far from unity. Therefore, taking into account both temperature limits, 
the analysis of Fig. \ref{cv2} of the behavior of the $C_V$ curve suggests that $\zeta$ must be in the range between the values $\zeta\approx \pm10^{43}$.

Let us notice that the upper boundary curve in Fig. \ref{zeta2} represents the case for which $\tilde C_V=0$ for all possible temperature ranges. It means that even for high temperatures, in principle, one could have a vanishing specific heat. 
Notice that there exists a maximum for around $y\approx0.48$ for which $\zeta\approx 2\times10^{45}$. Similarly as for $\xi=0$, it can be interpreted as the onset of a potential lack of a stable equilibrium configuration, providing to negative heat capacities happening in systems with long-range interactions.
The bottom boundary curve in Fig. \ref{zeta2} shows $\tilde C_V=1$ for which the temperatures can be high ($y\approx 1$): normal "classical" effect, but also as low as $y\approx 0.2$: leading to a nonphysical effect.

\section{(Anti-)Snyder-de Sitter and the generalized extended (GEUP) case}
As another possible extension of quantum mechanics one can consider a different type of noncommutative background, for example Snyder-de Sitter (SdS) model. SdS is a
 generalization of the Snyder model to a spacetime
background of constant curvature which includes both noncommutative coordinates and noncommutative momenta (in dimensions higher than 1). In the quantum phase space, in Heisenberg cross commutation relations, terms dependent on both positions $\hat{x}$ and momenta $\hat{p}$ appear. This model leads to a different type of the modification of the uncertainty principle, often referred to as generalized extended GEUP/EGUP, which admit both fundamental parameters $\zeta$ and $\alpha$. These parameters may be related to the Planck length and cosmological constant, respectively or may be related to the possible NC QM extension scales.

In the 1D case, the quantum mechanical phase space commutation relation for the (a)SdS model is:
\begin{equation}\label{1dpx_geup}
    \lbrack \hat{x},\hat{p}]=i\hbar \left( 1+\alpha\hat{x}^2+\zeta \hat{p}^2+\sqrt{\alpha \zeta} (\hat{x}\hat{p}+\hat{p}\hat{x})\right). 
\end{equation}
Representation of position and momentum operators $\hat{x},\hat{p}$ acting on a Hilbert space of functions (in momentum space) has been considered in \cite{Mignemi:2011wh}. Moreover, the exact solution of the quantum harmonic oscillator in the (a)SdS case \eqref{1dpx_geup} has also been obtained (although in the simplified case), therein. The energy levels of the harmonic oscillator are
\begin{eqnarray}
E_{n} &=&\hbar\omega \left[ \left(n + \frac{1}{2}\right)\sqrt{1 + B^2 } + \left(n^2+n+\frac{1}{2}\right)B\right]\nonumber\\
&&
\label{En1D_geup}
\end{eqnarray}
where $B=\frac{\hbar\mu(\zeta\omega+\alpha/\omega)}{2}$\footnote{Note that the corrections to the case of the standard quantum harmonic oscillator are of order  $\hbar\mu(\zeta\omega+\alpha/\omega)$ and that these correction exhibit a duality $\zeta\omega \longleftrightarrow \alpha/\omega $.}.

For $\alpha\to 0$ in $B$ in \eqref{En1D_geup}, we recover the energy levels from the previous section (i.e. $B\to A$ for the special realization choice with $\xi=1/2$ - the original Snyder realization) and we recover \eqref{En1D_27short} from \eqref{En1D_geup}. Taking $\zeta \to 0$ in \eqref{En1D_geup} we will get the extended uncertainty principle (EUP) case corresponding to 
\begin{eqnarray}
\left[\hat{x},\hat{p}\right] &=&i\hbar\left( 1+\zeta x^2\right)
\end{eqnarray}
appearing in (anti-)de Sitter models with commutative coordinates but noncommutative momenta (in dimensions higher than 1), see e.g. \cite{Pachol:2024hiz} for the discussion on various cases arising as limits of SdS model and the investigation on the density of states in all such cases.

Using \eqref{En1D_geup} to model $N$ harmonic oscillators in $3D$, in SdS background, we can use the results from the previous section. 
Therefore
\begin{align}
    Z&=\sum_{n=0}^\infty e^{-\frac{E_n}{k_BT}}
    =e^{-\frac{1}{2}\beta\hbar\omega (1+B)}\sum_{n=0}^\infty e^{-\beta\hbar\omega \left[n(1+B) + n^2B\right]}\nonumber\\
&
\label{sumB}
    =e^{-\beta E_0}\sum_{n=0}^\infty e^{-\beta\hbar\omega \left[n(1+B) + n^2B\right]},
\end{align}
which is convergent for $B>-\frac{1}{2}$, similarly as in the previous case.
In the unmodified limit $B\to 0$ (i.e. $\zeta\to 0$ and $\alpha\to 0)$ we recover the well-known canonical case consisting of the geometric series.
In the presence of noncommutative background (i.e. both nonzero $\zeta$ and $\alpha$) we consider $B$ as a small parameter and perform similar approximations as previously and obtain
\begin{equation}\label{Z_B}
     Z=e^{-\frac{\beta\hbar\omega}{2}}\Big(\frac{e^{\beta\hbar\omega}}{e^{\beta\hbar\omega}-1}-B\beta\hbar\omega (\frac{e^{\beta\hbar\omega}(e^{\beta\hbar\omega}+1)^2}{2(e^{\beta\hbar\omega}-1)^3}
     \Big)
\end{equation}
where the correction to the undeformed partition function is clear. The positivity of the partition function now is provided by ensuring 
     $\Big(1-B\beta\hbar\omega \frac{(e^{\beta\hbar\omega}+1)^2}{2(e^{\beta\hbar\omega}-1)^2}
     \Big)>0$.
Note, that there is no choice for different realizations in this case (i.e. we do not have different choices of $\xi$).

Considering different temperature regimes we find
\begin{equation}\label{signB}
B <
\begin{cases}
&\frac{2k_B T}{\hbar\omega},\qquad   k_B T<< \hbar\omega,\\
&\frac{\hbar\omega}{2k_B T},\qquad  k_B T>> \hbar\omega.
\end{cases}
\end{equation} 
We get the following frequency dependent inequality
\begin{equation}\label{signBB}
(\zeta+\frac{\alpha}{\omega^2}) <
\begin{cases}
&\frac{4k_B T}{\mu(\hbar\omega)^2},\qquad  k_B T<< \hbar\omega,\\
& \frac{1}{\mu k_B T},\qquad  k_B T>> \hbar\omega.
\end{cases}
\end{equation} 
Since the derivative of the partition function \eqref{Z_B} is with respect to $\beta=\frac{1}{k_B T}$ and the $\beta$ does not appear in $B$ then the result will be completely analogous to \eqref{u_total}
{\small 
\begin{align}\label{u_totalB}
    U    =& \frac{3N}{2}\hbar\omega\coth
   \left(\frac{\hbar\omega\beta}{2}\right)+\\ 
   +&\frac{3NB}{8} \hbar\omega\sinh\left(\hbar\omega\beta\right)\csch^4\left(\frac{\hbar\omega\beta}{2}\right)
   \left(\sinh \left(\hbar\omega\beta\right)-2\hbar\omega\beta\right).
   \nonumber
\end{align}
}

Considering the asymptotical behavior for both $\hbar\omega\gg k_B T$ as well as $\hbar\omega\ll k_B T$ one obtains (compare with \eqref{uf})
{\small 
\begin{equation}\label{ufB}
    U=
    \begin{cases}
	\frac{3}{2}N \hbar\omega  \left(
    1+B\left( 1-4\frac{\hbar \omega}{k_B T} e^{-\frac{\hbar \omega
   }{k_B T}}\right)\right), & k_B T \ll \hbar \omega,\\
          3Nk_BT \left(1-2B\frac{ k_B T}{\hbar\omega}\right) , & k_B T \gg \hbar \omega.
		 \end{cases}
\end{equation}
}
It is now straightforward to obtain the heat capacity of the considered system and we obtain (compare with \eqref{specific0}):
{\small 
\begin{align}\label{specificB}
    C_V=&\frac{dU}{dT}=\frac{3 N (\hbar\omega)^2 }{4 k_B
   T^2\text{sinh}^2\left(\frac{\hbar\omega}{2 k_B T}\right)}\\
   -&\frac{3B N \left(\frac{\hbar\omega}{k_BT}\right)^2  \left[\hbar \omega  (\cosh
   \left(\frac{\hbar\omega}{k_BT}\right)+2)-2 k_B T \sinh \left(\frac{\hbar\omega}{k_BT}\right)\right]}{4 T\text{sinh}^4\left(\frac{\hbar\omega}{2k_BT}\right) }.\nonumber
\end{align}
}
Calculating the critical values of $B_{crit}$ for which $C_V=0$ we obtain in this case (where $B_{crit}=\frac{1}{2}\mu \hbar \left( \omega \zeta_{crit} +\frac{\alpha_{crit} }{\omega }\right) $ is related with the critical values of NC parameters $\zeta_{crit}$ and $\alpha_{crit}$)
\footnote{in the same shortcut notation $y=\frac{1}{x}=\frac{k_{B}T}{\hbar \omega }$ and $x=\frac{\hbar\omega}{k_B T}$.}
\begin{equation}
    \label{critB2}
\left(  \zeta_{crit} +\frac{\alpha_{crit}}{\omega^2 }\right) =\frac{2{\sinh}^{2}\left( 
\frac{1}{2y}\right) }{\mu \hbar\omega \left( \frac{1}{y}(\cosh \left(\frac{1}{y}\right)
+2)-2\sinh\left( \frac{1}{y}\right) \right) }.
\end{equation} 
We note that there is a dependence on frequency on the left hand side. The approximations for low and high temperatures are as follows:
{\small 
\begin{equation}\label{C_VB}
    C_V=
    \begin{cases}
	3  N  k_B e^{-\frac{\hbar\omega}{k_B T}}(\frac{\hbar\omega}{k_BT})^2
(1-2 B\hbar\omega), & \text{if $k_B T \ll \hbar \omega$ }\\
          3 N k_B-12 k_B N B\frac{ k_B T}{\hbar \omega } , & \text{if $k_B T \gg \hbar \omega$ },
		 \end{cases}
\end{equation}
}
 Using \eqref{critB2} and values for $\mu$ and $\hbar\omega$ for a diamond (footnote 2), we obtain: 
 \begin{equation}
    \label{critB3}
\left(  \zeta_{crit} +\frac{\alpha_{crit}}{\omega^2 }\right) =\frac{ 6.596\times 10^{45}{\sinh}^{2}\left( 
\frac{1}{2y}\right) }{\left( \frac{1}{y}(\cosh \left(\frac{1}{y}\right)
+2)-2\sinh\left( \frac{1}{y}\right) \right) }.
\end{equation} 
(cf. \eqref{zeta_crit}).
The discussion presented in the previous section in the case of realization $\xi=\frac{1}{2}$ gives similar bound for the sum of the NC parameters $\left(  \zeta +\frac{\alpha}{\omega^2 }\right)$.
 
\section{Conclusions}
The aim of this paper was to propose an investigation into how low-energy consequences of modified quantum phase space models could be identified in table-top experiments. Such models are often linked to phenomenological effects of quantum theories of gravity via e.g. GUP or GEUP models.

We focused on deriving noncommutative corrections to crystalline solids within the simplified Einstein model and on assessing their impact on physically observable quantities, such as the specific heat.
We find that these corrections may lead to nonphysical behavior of the crystal under consideration, giving rise to certain bounds imposed on the correction parameters. In particular, the partition function acquires corrections that can, in principle, render it negative, which already allows us to place constraints on the parameters of the modified quantum phase space models (or parameters in the modified uncertainty principles). 

Furthermore, we derived corrections to physical observables such as internal energy and specific heat. However, for general values of the noncommutative parameters, these modifications may lead to large deviations from their well-established properties.
For example, the specific heat can become negative or its normalized value can exceed unity at high temperatures 
 ($k_B T \propto \hbar \omega$). We have taken those physical constraints into account in order to find bounds on the parameters of the models.
 All of the analysis was performed for a diamond, which is well described, at low temperatures regime, by the Einstein model of crystalline solids.

An interesting feature is illustrated in Figs. \ref{zeta1} and \ref{zeta2}. The theory parameter appears to be sensitive to the energy/temperature regime in which the thermodynamical property is probed \cite{Lope-Oter:2023urz,Pachol:2023tqa,wojnar2026waves}. In particular, stronger constraints are obtained at higher temperatures, whereas in the limit $T\rightarrow0$ the parameter becomes effectively unconstrained, allowing for arbitrarily large absolute values.

One possible explanation is that, at higher temperatures, the increased vibrational activity of particles enhances the role of momentum-dependent corrections in the uncertainty relation. Alternatively, this behavior may be understood in terms of a deformation of phase-space cells, which becomes more significant at higher momenta, reflecting the momentum dependence of the effective phase-space volume available to quantum states. 
Therefore, in order to remain consistent with the well-established physics of crystals at accessible temperatures (noting that the Einstein model is not adequate for describing solids at high temperatures), the correction term must be subject to stronger suppression.

\section*{Acknowledgements}
This article is based upon work from COST Action FuSe CA24101 and COST Action BridgeQG CA23130, supported by COST (European Cooperation in Science and Technology).\\ A.P. acknowledges the support of the Polish NCN Grant No.
2022/45/B/ST2/01067.\\
Part of
this work was conducted while A.P. was visiting the Okinawa Institute of Science and Technology (OIST) through the Theoretical Sciences Visiting Program (TSVP). A.P. would like to thank the
TSVP for enabling her
visit, and for the hospitality and excellent working conditions during her time there.
\appendix 
\section{Constraints from the positivity of the partition function}
\label{ln_ineq}
Let us introduce a dimensionless notation with $x=\beta\hbar\omega=\frac{\hbar\omega}{k_B T}$ in the inequality \eqref{signA}:
$$
   \Big(1-A\beta\hbar\omega \frac{(e^{\beta\hbar\omega}+1)^2}{2(e^{\beta\hbar\omega}-1)^2}
     \Big)>0 .$$
Since $x$ and both brackets including its exponents of $x$ are always positive for any temperature, one ends up with the following relation:
\begin{equation}\label{signAA}
A <\frac{2(e^{x}-1)^2 }{x(e^{x}+1)^2}.
\end{equation}
Considering different temperature regimes, we obtain:
\begin{equation}\label{signAAA}
A <
\begin{cases}
&\frac{2}{x}=\frac{2k_B T}{\hbar\omega},\qquad\quad\quad\quad\quad\quad\mbox{for }  k_B T<< \hbar\omega,\\
&\frac{1}{2}x
-\frac{1}{12}x^3\approx\frac{1}{2}\frac{\hbar\omega}{k_B T},\qquad\mbox{ for } k_B T>> \hbar\omega.
\end{cases}
\end{equation} 
Since $A=\frac{\mu}{2}\hbar\omega\theta=\frac{\mu}{2}\hbar\omega\zeta(3\xi-\frac{1}{2})$, we get the following mass dependent inequality (mass $\mu$ is always positive):
\begin{equation}\label{signAAAA}
\mu\zeta(3\xi-\frac{1}{2}) <
\begin{cases}
&\frac{4k_B T}{(\hbar\omega)^2},\qquad  k_B T<< \hbar\omega\\
& \frac{1}{k_B T},\qquad  k_B T>> \hbar\omega.
\end{cases}
\end{equation} 
In order to get some numerical values for the parameter $\zeta$, let us consider a diamond crystal (see footnote 2) which is well described by the Einstein model in low temperatures. 

Note that the case with $\xi=0$ (i.e. the Maggiore realization of the Snyder model) automatically satisfies this inequality for any temperature regime with the noncommutative parameter $\zeta$ being  positive.
\begin{figure}[ht]
     \includegraphics[scale=0.43]{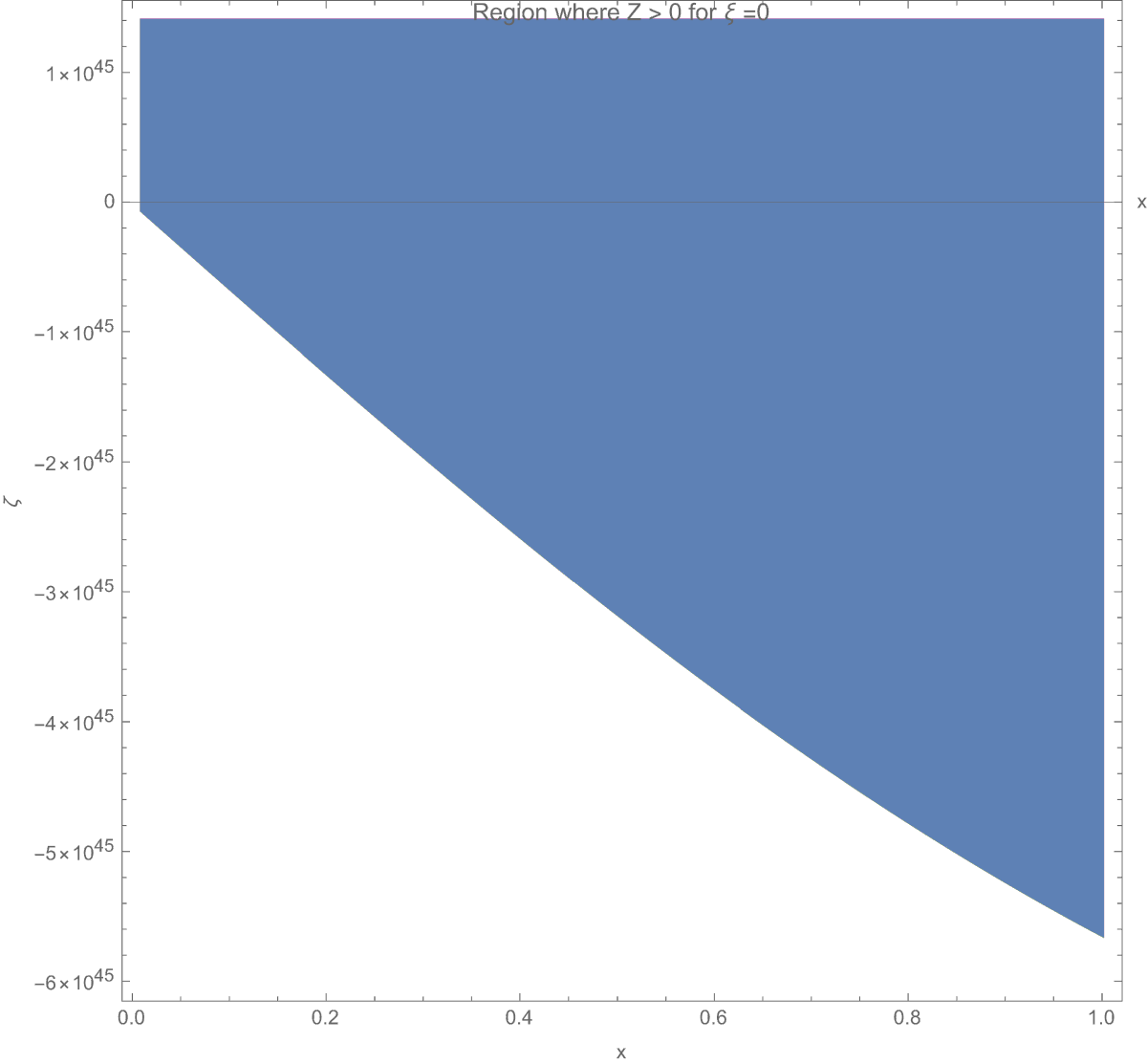}
     \caption{Allowed values of the parameter $\zeta$ in different temperature range ($x=\frac{\hbar\omega}{k_B T}$) for which the argument of $\ln$ in the partition function, i.e. \eqref{signA} is always positive (shaded regions) for the case $\xi=0$. Here, as before, $\mu=9.96\times 10^{-27}$kg for a diamond crystal and we took $\hbar\omega=3.04\times10^{-20}$J for a phonon energy.}
     \label{dozwolone1}
\end{figure}
\begin{figure}[!h]
     \includegraphics[scale=0.43]{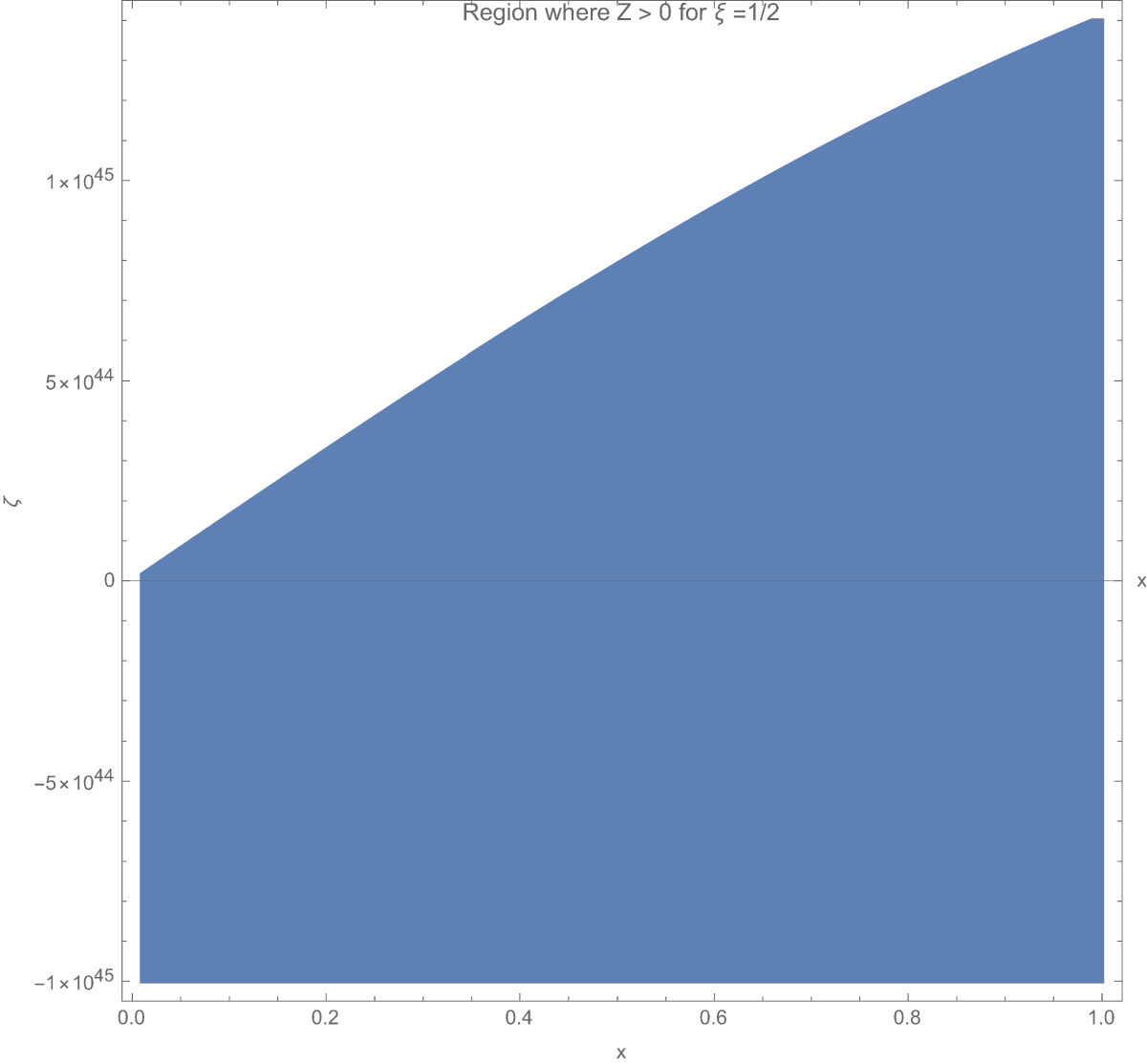}
     \caption{Allowed values of the parameter $\zeta$ in different temperature range ($x=\frac{\hbar\omega}{k_B T}$) for which the partition function \eqref{signA} is always positive (shaded regions) for the case $\xi=\frac{1}{2}$. Here, as before, $\mu=9.96\times 10^{-27}$kg for a diamond crystal and we took $\hbar\omega=3.04\times10^{-20}$J for a phonon energy.}
     \label{dozwolone2}
\end{figure}

However, we may also consider, depending on the temperature regime, $\zeta$ to take negative values so that the above inequalities are satisfied. In this case one deals with the anti-Snyder model \cite{Mignemi:2011gr}, especially in the high temperature regime, see Fig. \ref{dozwolone1}. 
Analyzing \eqref{signAAAA} with $\xi=0$ we have
\begin{equation}
\zeta >
\begin{cases}
&-\frac{8k_B T}{\mu(\hbar\omega)^2},\qquad  k_B T<< \hbar\omega\\
& -\frac{2}{\mu k_B T},\qquad  k_B T>> \hbar\omega.
\end{cases}
\end{equation} 
On the other hand, for the case $\xi=\frac{1}{2}$ (the original Snyder realization) the region of positive values $\zeta$ that meets this constraint is given in Fig. \ref{dozwolone2}, while the negative values of $\zeta$ in this case would automatically satisfy the inequality \eqref{signAAAA}.

\end{document}